\documentclass[useAMS, usenatbib, usegraphicx, fleqn]{mn2e}
\usepackage{times, amsmath, amssymb, bm}
\usepackage[breaklinks, colorlinks, citecolor=blue, linkcolor=black, urlcolor=black]{hyperref}

\newcommand{\uv}[1]{\ensuremath{\bm{\hat{#1}}}}
\newcommand{\wj}[6]{
\left(
\begin{array}{ccc}
\! #1\! & #2\! & #3\! \\
\! #4\! & #5\! & #6\!
\end{array}
\right)}

\title[Multiplicative errors in the galaxy power spectrum]{Multiplicative errors in the galaxy power spectrum: self-calibration of unknown photometric systematics for precision cosmology}

\author[D.~L.~Shafer and D.~Huterer]{Daniel~L.~Shafer\thanks{E-mail: dlshafer@umich.edu (DLS); huterer@umich.edu (DH)} and Dragan~Huterer\footnotemark[1]\\
Department of Physics, University of Michigan, 450 Church Street, Ann Arbor, MI 48109-1040, USA}

\pagerange{\pageref{firstpage}--\pageref{lastpage}} \pubyear{2014}

\begin{document}
\label{firstpage}
\maketitle

\begin{abstract}
We develop a general method to ``self-calibrate'' observations of galaxy clustering with respect to systematics associated with photometric calibration errors. We first point out the danger posed by the multiplicative effect of calibration errors, where large-angle error propagates to small scales and may be significant even if the large-scale information is cleaned or not used in the cosmological analysis. We then propose a method to measure the arbitrary large-scale calibration errors and use these measurements to correct the small-scale (high-multipole) power which is most useful for constraining the majority of cosmological parameters. We demonstrate the effectiveness of our approach on synthetic examples and briefly discuss how it may be applied to real data.
\end{abstract}

\begin{keywords}
large-scale structure of Universe -- cosmological parameters -- dark energy -- galaxies: statistics -- techniques: photometric
\end{keywords}

\section{Introduction}
Observations of large-scale structure (LSS) have proven to be a powerful probe of cosmology in recent years. Earlier large galaxy surveys, such as CfA \citep{deLapparent:1985ta}, APM \citep{Maddox:1990yw}, and 2dF \citep{Colless:2001gk, Cole:2005sx}, have paved the way for modern surveys like WiggleZ \citep{Parkinson:2012vd}, SDSS \citep{York:2000gk, Tegmark:2003uf, Abazajian:2008wr}, and the Baryon Oscillation Spectroscopic Survey (BOSS), the current incarnation of SDSS \citep{Ahn:2012fh, Dawson:2012va}. These surveys have identified millions of galaxies and obtained spectra (and therefore redshifts) of over one million, allowing us to map the three-dimensional distribution of matter in the Universe. The enormous data sets resulting from these surveys are often distilled into measurements of the location of the peak in the correlation function corresponding to the scale of baryon acoustic oscillations (BAO), the imprint on structure resulting from sound waves propagating in the early Universe \citep{Eisenstein:2005su, Percival:2009xn, Padmanabhan:2012hf, Anderson:2012sa, Ross:2014qpa}. These BAO measurements, some of which are now at the percent level \citep{Anderson:2013zyy}, have been crucial in that they complement other probes of cosmic expansion to help break degeneracies between key cosmological parameters.

But there is much more information in the power spectrum than just the primary BAO peak, and with ongoing surveys like the Dark Energy Survey (DES; \citealt{Abbott:2005bi}), which is on track to identify $\sim$300 million galaxies, the galaxy power spectrum as a whole will complement other probes of cosmology (CMB power spectra, Type Ia supernovae, weak lensing, etc.) to place tight constraints on dark energy and other cosmological parameters. Planning is already underway for future wide-field LSS surveys, such as LSST \citep{Ivezic:2008fe} and \textit{Euclid} \citep{Laureijs:2011gra}.

Making these observations suitable for cosmology is not trivial. With the enormous statistical power of surveys like DES, control of systematics becomes crucial, especially at small scales where cosmic variance is small. A major class of systematic errors is \emph{photometric calibration} errors, by which we mean any systematic that effectively causes the magnitude limit of the sample to vary across the sky, thus biasing the true galaxy power spectrum. A number of recent observations \citep{Goto:2012yc, Pullen:2012rd, Ho:2012vy, Ho:2013lda, Agarwal:2013qta, Giannantonio:2013uqa, Agarwal:2013ajb} show a significant excess of power at large scales that likely results from such calibration errors that have not been accounted for. Recent work \citep{Ross:2011cz, Pullen:2012rd, Ho:2012vy, Agarwal:2013ajb, Leistedt:2013gfa, Leistedt:2014wia} has focused on mitigating these systematics in order to probe the underlying cosmology.

\cite{Huterer:2012zs}, which we will refer to as H13, introduced a formalism for quantifying the effect of an arbitrary photometric calibration error. They found that in order to use information from large scales to constrain cosmological parameters, the root-mean-square variation due to the calibration field must be $\sim$0.001--0.01 mag or less in order to avoid significantly biasing cosmological parameters. This is a very stringent requirement.

A potentially more dangerous effect is the \textit{multiplicative} leakage of these large-angle errors to small angular scales, where most cosmological information resides. As discussed in H13, the observed number density of galaxies is given by $N_\text{obs}(\uv{n}) = \left[1 + c(\uv{n}) \right] N(\uv{n})$, where the true number density $N(\uv{n})$ is modulated by a calibration-error field $c(\uv{n})$, which is directly related to a host of interrelated photometric effects (survey depth, completeness, atmospheric conditions, galactic dust, etc.). Even though the calibration error is significant only at large angular scales, it multiplies the true galaxy field, so the observed field is affected on \emph{all} angular scales. This multiplicative effect, to our knowledge first pointed out in H13 in the context of LSS systematics, is further studied in this work.

Current methods to clean the power spectra have been impressively efficient, but they typically rely on systematics templates -- prior knowledge of the relative spatial variation of the contamination across the sky due to a known systematic. The methods rely on the assumption that these templates are correct and that the set of templates is complete; any \emph{unknown} large-scale systematic that is not covered by one of the templates (or some combination thereof) will not be accounted for.

Mode projection (or extended mode projection, see \citealt{Leistedt:2014wia}) is a particularly effective method. Essentially, it is a way of marginalizing over spatially varying patterns on the sky that are expected to be caused by various systematics. While mode projection has been shown to be effective at mitigating the added power from known systematics, this method cannot remove multiplicative errors. To understand why, suppose for simplicity that a single systematic effect modulates the observed galaxy densities and that the shape of the template is that of a pure spherical harmonic (so that $c(\uv{n}) \propto Y_{\ell m}(\uv{n})$ for some $\ell$, $m$). This modulation will then not only add power at the angular scale $\ell$, but it will also affect the power at \textit{all} other multipoles. The obvious way to ``project out'' this mode, at least in principle, is to simply ignore that one contaminated $m$ mode when estimating the variance $C_\ell$. The additive error is removed entirely with little loss of cosmological information, but other multipoles have still been affected by the multiplicative effect in accordance with our Eq.~\eqref{eq:Tl} below.

In this paper, we study a new approach that is both alternative and complementary to previously employed techniques: using some of the power spectrum observations themselves to directly measure the systematic contamination and correct the rest of the measurements for a cosmological analysis. Since the calibration errors are expected to enter at large scales and then fall off quickly at higher multipoles (smaller scales), one may interpret the low-multipole power spectrum as measurements of the systematics and use these to correct the power spectrum at high multipoles, sacrificing some cosmological information from large scales to remove the multiplicative error and obtain unbiased estimates of cosmological parameters from small scales. The benefit of this approach is that no templates or otherwise detailed modelling of the systematics is required at this level.

The rest of the paper is organized as follows. In Sec.~\ref{sec:method}, we review and extend the calibration formalism introduced in H13, discuss our Fisher matrix formalism, and describe a fiducial model and DES-like survey. In Sec.~\ref{sec:results}, we quantify the effect of multiplicative calibration error for our fiducial survey and demonstrate the self-calibration method. In Sec.~\ref{sec:discuss}, we summarize our conclusions and discuss how one might apply the self-calibration method to real data.

\section{Methodology} \label{sec:method}
In this section, we outline our formalism to describe the calibration errors, review and extend the Fisher matrix for the galaxy power spectrum, and detail our fiducial model and survey.

\subsection{Calibration Error Formalism} \label{sec:formalism}

In the absence of all systematics, we would observe the true number density of galaxies on the sky, which we expand in spherical harmonics:
\begin{equation}
\delta(\uv{n}) \equiv \frac{N(\uv{n}) - \bar{N}}{\bar{N}} = \sum_{\ell=0}^{\infty} \sum_{m=-\ell}^{\ell} a_{\ell m} Y_{\ell m}(\uv{n}) \ ,
\label{eq:delta}
\end{equation}
where a bar denotes a sky average and where the monopole vanishes ($a_{0 0} = 0$) since it is proportional to the average overdensity on the sky, which is zero by construction. The coefficients $a_{\ell m}$ are expected to be Gaussian random variables with a mean of zero and a variance that depends on the cosmological model. The various $m$ modes are statistically independent under the assumption of isotropy, so we have the familiar relations
\begin{align}
\langle a_{\ell m} a_{\ell' m'}^* \rangle &= \delta_{\ell \ell'} \, \delta_{m m'} \, C_\ell \ , \\
\langle a_{\ell m} \rangle &= 0 \ .
\end{align}
Following H13, we now consider the effect of an arbitrary variation in the limiting magnitude $\delta m_\text{max}(\uv{n})$ of the photometric survey due to calibration variation across the sky. This magnitude variation implies a \textit{relative} variation in galaxy counts $[\delta N/N](\uv{n}) \propto \delta m_\text{max}(\uv{n})$, where the constant of proportionality depends on the faint-end slope of the luminosity function $s(z) \equiv d \log_{10} N(z, m) / dm \vert_{m_\text{max}}$ that may depend on redshift but does not depend on direction. More generally, one can write $[\delta N/N](\uv{n}) \equiv c(\uv{n})$ so that the observed galaxy number density $N_\text{obs}(\uv{n})$ is equal to the true number density $N(\uv{n})$ modulated by the field $c(\uv{n})$, which we can also expand in spherical harmonics:
\begin{align}
N_\text{obs}(\uv{n}) &= \left[1 + c(\uv{n}) \right] N(\uv{n}) \ , \\
c(\uv{n}) &= \sum_{\ell m} c_{\ell m} Y_{\ell m}(\uv{n}) \ .
\end{align}
The calibration coefficients $c_{\ell m}$ are deterministic (not inherently stochastic) and thus have the trivial statistical properties
\begin{align}
\langle c_{\ell m} c_{\ell' m'}^* \rangle &= c_{\ell m} c_{\ell' m'}^* \ , \\
\langle c_{\ell m} \rangle &= c_{\ell m} \ .
\end{align}
In other words, the calibration-error field is a fixed pattern on the sky, and there is no loss of generality in making this assumption to simplify the analysis.  The observed overdensity is given by
\begin{align}
\delta_\text{obs}(\uv{n}) &\equiv \frac{N_\text{obs}(\uv{n}) - \bar{N}_\text{obs}}{\bar{N}_\text{obs}} = \frac{\left[1 + c(\uv{n}) \right] N(\uv{n})}{\bar{N} (1 + \epsilon)} - 1 \nonumber \\
&= \frac{1}{1 + \epsilon} \left[\delta(\uv{n}) + c(\uv{n}) + c(\uv{n}) \delta(\uv{n}) - \epsilon \right] \ ,
\end{align}
where $\bar{N}_\text{obs} = \bar{N}(1 + \epsilon)$ is the observed mean number of galaxies per pixel on the sky. Then $\epsilon$ is defined by
\begin{align}
\epsilon &\equiv \frac{1}{\bar{N}} \ \overline{c(\uv{n}) N(\uv{n})} = \overline{c(\uv{n}) \left[1 + \delta(\uv{n}) \right]} \nonumber \\
&= \frac{c_{0 0}}{\sqrt{4\pi}} + \frac{1}{4 \pi} \sum_{\ell m} c_{\ell m} a_{\ell m}^* \ , \label{eq:epsilon}
\end{align}
where the overbar again denotes a sky average.

Expanding the observed overdensity in spherical harmonics,
\begin{equation}
\delta_\text{obs}(\uv{n}) = \sum_{\ell m} t_{\ell m} Y_{\ell m}(\uv{n}) \ ,
\end{equation}
we calculate the expansion coefficients of $\delta_\text{obs}$ to be
\begin{align}
t_{\ell m} = \frac{1}{1 + \epsilon} &\left[a_{\ell m} + c_{\ell m} - \sqrt{4\pi} \ \epsilon \ \delta_{\ell 0} \vphantom{\sum_{\substack{\ell_1 m_1 \\ \ell_2 m_2}}} \right. \\[-0.7cm]
&\qquad \qquad \quad + \left. \sum_{\substack{\ell_1 m_1 \\ \ell_2 m_2}} R^{\ell_1 \, \ell_2 \, \ell}_{m_1 m_2 m} \, c_{\ell_1 m_1} a_{\ell_2 m_2} \right] \ , \nonumber
\end{align}
where the coupling coefficient $R^{\ell_1 \, \ell_2 \, \ell}_{m_1 m_2 m}$ is a variant of the Gaunt coefficient, the result of integrating a product of three spherical harmonics. It can be written in terms of Wigner~3-$j$ symbols:
\begin{align}
R^{\ell_1 \, \ell_2 \, \ell}_{m_1 m_2 m} \equiv (-1)^m &\sqrt{\frac{(2\ell_1 + 1)(2\ell_2 + 1)(2\ell + 1)}{4\pi}} \label{eq:R} \\
&\times \wj{\ell_1}{\ell_2}{\ell}{0}{0}{0} \wj{\ell_1}{\ell_2}{\ell}{m_1}{m_2}{-m} \ . \nonumber
\end{align}

In this formalism, how will the observed power be related to the true power $C_\ell$? If the observed coefficients are $t_{\ell m}$, then the observed (potentially biased) power is given by (Appendix~\ref{sec:derive})
\begin{align}
T_\ell &= \frac{\sum_m \langle |t_{\ell m}|^2 \rangle}{2\ell + 1} \nonumber \\
&= C_\ell + C_\ell^\text{cal} - \frac{1}{2\pi} C_\ell^\text{cal} C_\ell \label{eq:Tl} \\
&\quad + \frac{1}{4\pi} \sum_{\ell_1 \ell_2} (2\ell_1 + 1) \, C_{\ell_1}^\text{cal} \wj{\ell_1}{\ell_2}{\ell}{0}{0}{0}^2 (2\ell_2 + 1) \, C_{\ell_2} \ , \nonumber
\end{align}
where we have defined the calibration-error contribution to the observed power as
\begin{equation}
C_\ell^\text{cal} \equiv \sum_m |c_{\ell m}|^2/(2\ell + 1) \ .
\label{eq:Clcal}
\end{equation}
It is worth making several comments about Eq.~\eqref{eq:Tl}, where the observed angular power spectrum $T_\ell$ is given as a function of the true power $C_\ell$, the calibration-error power $C_\ell^\text{cal}$, and various geometric coupling factors. The second term on the right-hand side, $C_\ell^\text{cal}$, represents the \emph{additive} effect of calibration error, which is typically important at large angular scales (such as $\ell \lesssim 20$). The other terms are products of the true and calibration-error power; they represent the \emph{multiplicative} effect of calibration errors that affects the observed power $T_\ell$ at all angular scales. Note that Eq.~\eqref{eq:Tl} is equivalent to the formula for $T_\ell$ from H13, though here we have substantially simplified it by defining away the $1/(1 + \epsilon)$ factor and using Wigner 3-$j$ relations (see Appendix~\ref{sec:derive} for more details).

Finally, we can gain some insight into the multiplicative contribution to $T_\ell$ by simplifying Eq.~\eqref{eq:Tl} under the assumption that $C_\ell^\text{cal}$ vanishes for multipoles greater than some cutoff multipole $\ell = \ell_\text{max, cal}$ and with the approximation that the true power $C_\ell$ is constant in the multipole range $\ell \pm \ell_\text{max, cal}$ that contributes to the sum over $\ell_2$. This allows us to factor $C_{\ell_2}$ out of the sum and apply the Wigner~3-$j$ relation Eq.~\eqref{eq:w3j2}. Then Eq.~\eqref{eq:Tl} simply becomes
\begin{equation}
T_\ell \simeq C_\ell + C_\ell^\text{cal} - \frac{1}{2\pi} C_\ell^\text{cal} C_\ell + \sigma^2_c \, C_\ell \ ,
\label{eq:Tlapprox}
\end{equation}
where $\sigma^2_c \equiv \text{Var}[c(\uv{n})] = \sum_{\ell = 1}^\infty (2\ell + 1) C_\ell^\text{cal}/(4\pi)$ is the variance of the calibration field across the sky. This is generally an excellent approximation, and it shows that the multiplicative effect (the last term) is roughly independent of the shape of the calibration power spectrum. Also, all $C_\ell$ at $\ell > \ell_\text{max, cal}$ are multiplied by roughly the same factor, an effect mimicking that of an incorrect galaxy bias.

\subsection{Fisher matrix and bias} \label{sec:fisher}

We now review and extend the standard LSS Fisher matrix formalism in order to forecast both the extent to which multiplicative calibration errors bias cosmological parameters and the uncertainty that results from trying to measure the contamination itself.

In the absence of systematics, and when cross-power spectra are assumed to vanish ($C_\ell^\text{ij} \equiv 0$ for $i \neq j$), the observables $C_\ell^{ii}$ are uncorrelated with a variance due to cosmic sampling variance and shot noise:
\begin{equation}
\text{Var}[C_\ell^{ii}] = \frac{2}{(2 \ell + 1) f_\text{sky}} \left(C_\ell^{ii} + \frac{1}{N^{i}} \right)^2 ,
\label{eq:cosmicvar}
\end{equation}
where $N^{i}$ is the number of galaxies per steradian for redshift bin $i$. Then the well-known Fisher matrix for measurements of the power spectrum is given by
\begin{equation}
F_{\alpha \beta} = \sum_i \sum_\ell \frac{\partial C_\ell^{ii}}{\partial p_\alpha} \frac{1}{\text{Var}[C_\ell^{ii}]} \frac{\partial C_{\ell}^{ii}}{\partial p_\beta} \ ,
\label{eq:fisher}
\end{equation}
where the $p_i$ are cosmological parameters.

We will be particularly interested in the effect of the presence of uncorrected-for calibration errors on cosmological parameters. A useful Fisher-matrix-based formalism is available to evaluate these effects \citep{Knox:1998fp, Huterer:2001yu}. Given a bias $\delta \mathbf{m}$ in a vector of observables $\mathbf{m}$ with covariance matrix $\mathbf{C}$, the linear estimate for the bias in cosmological parameters is
\begin{equation}
\delta \mathbf{p} = \mathbf{F}^{-1} \mathbf{D} \ \mathbf{C}^{-1} \delta \mathbf{m} \ ,
\end{equation}
where the matrix $\mathbf{D}$ contains the derivatives of the observables with respect to the parameters evaluated at their fiducial values: $D_{ij} = \partial m_j / \partial p_i$.

We can extend the Fisher matrix to measure both cosmological parameters and calibration-error parameters $C_\ell^\text{cal}$ from the observed (biased) angular power spectrum. Our observables are now the set of $T_\ell^{ii}$, and we calculate their covariance from Eq.~\eqref{eq:Tlapprox}. To a good approximation, they are uncorrelated with a variance
\begin{equation}
\text{Var}[T_\ell] \simeq \text{Var}[C_\ell] \left(1 + \sigma^2_c - \frac{1}{2\pi} C_\ell^\text{cal} \right)^2 ,
\end{equation}
where $\text{Var}[C_\ell]$ is the usual error from cosmic variance plus shot noise, as in Eq.~\eqref{eq:cosmicvar}, and the redshift bin indices have been suppressed. The derivatives with respect to the parameters are
\begin{align}
\frac{\partial T_\ell}{\partial p_i} &\simeq \frac{\partial C_\ell}{\partial p_i} \left(1 + \sigma^2_c - \frac{1}{2\pi} C_\ell^\text{cal} \right) \ , \\
\frac{\partial T_\ell}{\partial C_{\ell'}^\text{cal}} &\simeq \delta_{\ell \ell'} \left(1 - \frac{1}{2\pi} C_\ell \right) + \frac{2\ell' + 1}{4\pi} \, C_\ell \ .
\end{align}
Note that when none of the $C_\ell^\text{cal}$ are added as parameters, the Fisher matrix for cosmological parameters reduces to the usual Fisher matrix Eq.~\eqref{eq:fisher}, independent of the fiducial size of the calibration errors.

\subsection{Fiducial model and survey} \label{sec:fid}

At small scales, we use the Limber approximation, which ignores the contribution of radial modes, and model the angular power spectra of galaxy density fluctuations as
\begin{equation}
C_\ell^{ij} = b^i b^j \int_0^\infty \frac{H(z)}{r^2(z)} \, P\left(\frac{\ell + \frac{1}{2}}{r(z)}, z \right) \, W^i(z) \, W^j(z) \, dz \, ,
\end{equation}
where $b^i$ is the galaxy bias for the $i^\text{th}$ redshift bin (which we assume to be a constant), $H(z)$ is the Hubble parameter, $r(z)$ is the comoving distance, $P(k, z)$ is the power spectrum, and the weights are given by
\begin{equation}
W^i(z) = \frac{n(z)}{N^i} \left[\theta\left(z - z_\text{min}^i \right) - \theta\left(z - z_\text{max}^i \right) \right],
\end{equation}
where $\theta(x)$ is the Heaviside step function, $z_\text{min}^i$ and $z_\text{max}^i$ are the lower and upper bound of the $i^\text{th}$ redshift bin, and $n(z)$ is the radial distribution of galaxies per steradian. We use the transfer functions and nonlinear modelling of \textsc{camb} \citep{Lewis:1999bs} to compute and evolve the power spectrum.

The Limber approximation is not valid at the largest scales, so for $\ell \leq 30$, we use the full expression for the power spectrum:
\begin{align}
C_\ell^{ij} &= \frac{2}{\pi} \int_0^\infty P(k, 0) \, I_\ell^i(k) \, I_\ell^j(k) \, k^2 dk \, , \\
I_\ell^i(k) &\equiv  b^i \int_0^\infty W^i(z) \, D(z) \, j_\ell(k \, r(z)) \, dz \, ,
\end{align}
where $D(z)$ is the linear growth factor relative to $z = 0$ (since we are safely in the linear regime) and $j_\ell(x)$ is the spherical Bessel function of order $\ell$. Note that in the above we are assuming a flat universe.

\begin{figure}
\includegraphics[width=0.45\textwidth]{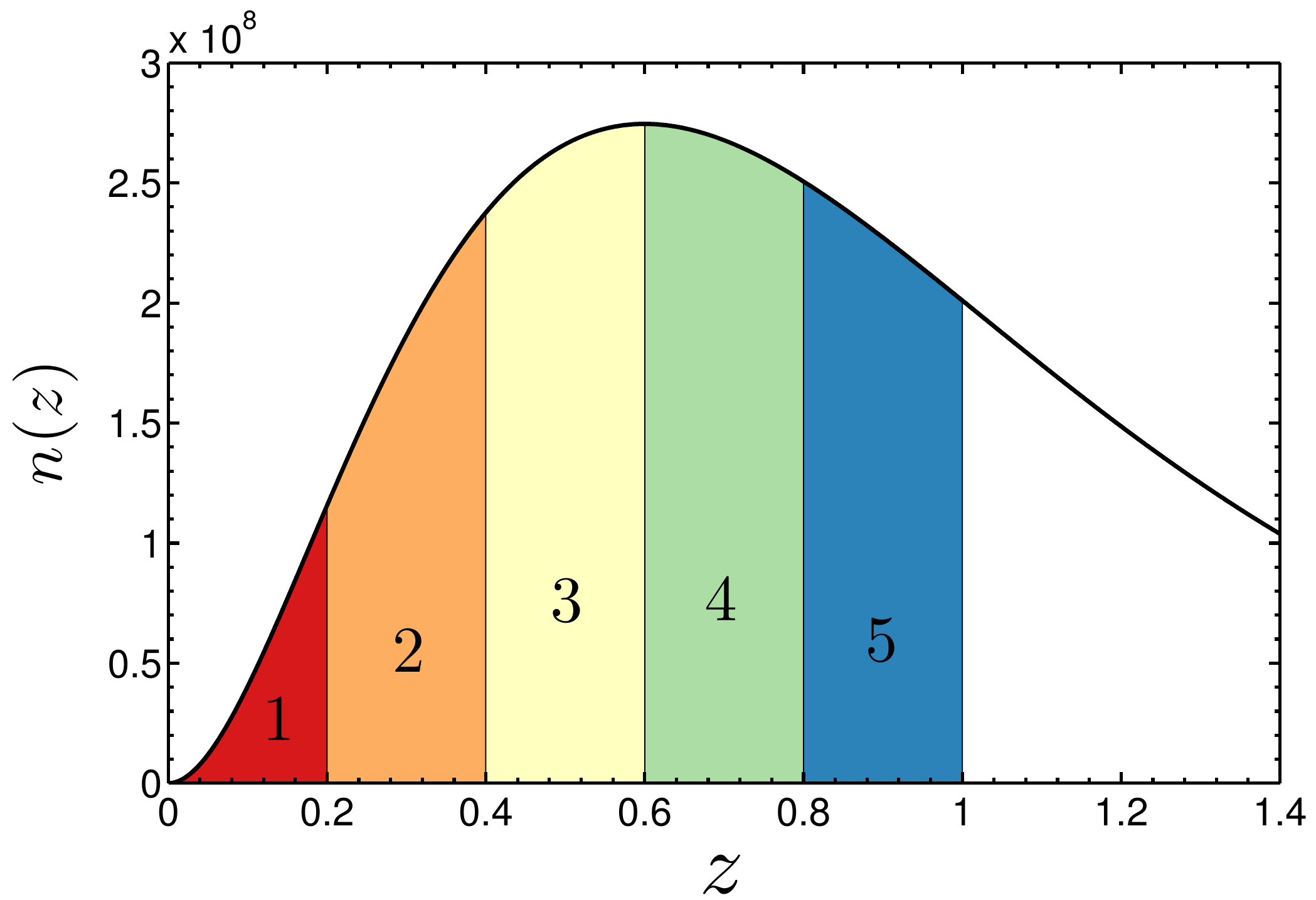}
\caption{Number density of galaxies per steradian for our fiducial survey. Galaxies are assigned to the five redshift bins in proportion to the areas of the coloured regions, each spanning $\Delta z = 0.2$.}
\label{fig:nz}
\end{figure}

Our fiducial DES-like survey covers 5,000~deg$^2$ (corresponding to $f_\text{sky} \simeq 0.12$) and identifies a total of 300~million galaxies ($\simeq$17 galaxies per arcmin$^2$). We split the sample into five tomographic redshift bins of width $\Delta z = 0.2$, centred at ${z = 0.1,~0.3,~0.5,~0.7,~\text{and}~0.9}$. We take the radial distribution of galaxies to be $n(z) \propto z^2 \exp(-z/z_0)$, with $z_0 = 0.3$, and divide the total number of galaxies among the redshift bins accordingly. As shown in Fig.~\ref{fig:nz}, the distribution  peaks at $z = 2 z_0 = 0.6$. We assume that the photometric redshifts can be determined well enough that the cross-power spectra between these bins are small enough to be ignored, making the power spectra for different $z$ slices statistically independent.

We choose for our fiducial cosmology a flat $\Lambda\text{CDM}$ model with $\Omega_m = 0.3$, $\Omega_m h^2 = 0.143$, $\Omega_b h^2 = 0.0222$, $n_s = 0.96$, and $10^9 A_s = 2.2$ for $k_0 = 0.05~\text{Mpc}^{-1}$, values which agree well with data from \textit{Planck} \citep{Ade:2013zuv} and other probes.

In our analysis, we allow the dark energy equation of state to vary along with the parameters above, though we keep $\Omega_m h^2$ and $\Omega_b h^2$ fixed at their fiducial values. In practice, Planck CMB measurements constrain these parameters very well (to $\sim$1\%), so we are effectively adding Planck priors to all of our constraints and assuming the remaining uncertainty to be negligible, which should be a reasonable approximation. We assume a (constant) galaxy bias $b^i = 2.2$ (same for all redshift bins), which we hold fixed for simplicity. In a full analysis, one should parametrize the bias appropriately and marginalize, since the bias parameters may have significant uncertainties and be somewhat degenerate with cosmological parameters. Allowing the dark energy equation of state to vary with time as $w(a) = w_0 + w_a (1 - a)$ \citep{Linder:2002et}, our fiducial parameter space is therefore five-dimensional ($\Omega_m$, $w_0$, $w_a$, $n_s$, $A_s$), though we briefly consider a constant equation of state (with fixed $w_a = 0$) as well.

\begin{figure}
\includegraphics[width=0.45\textwidth]{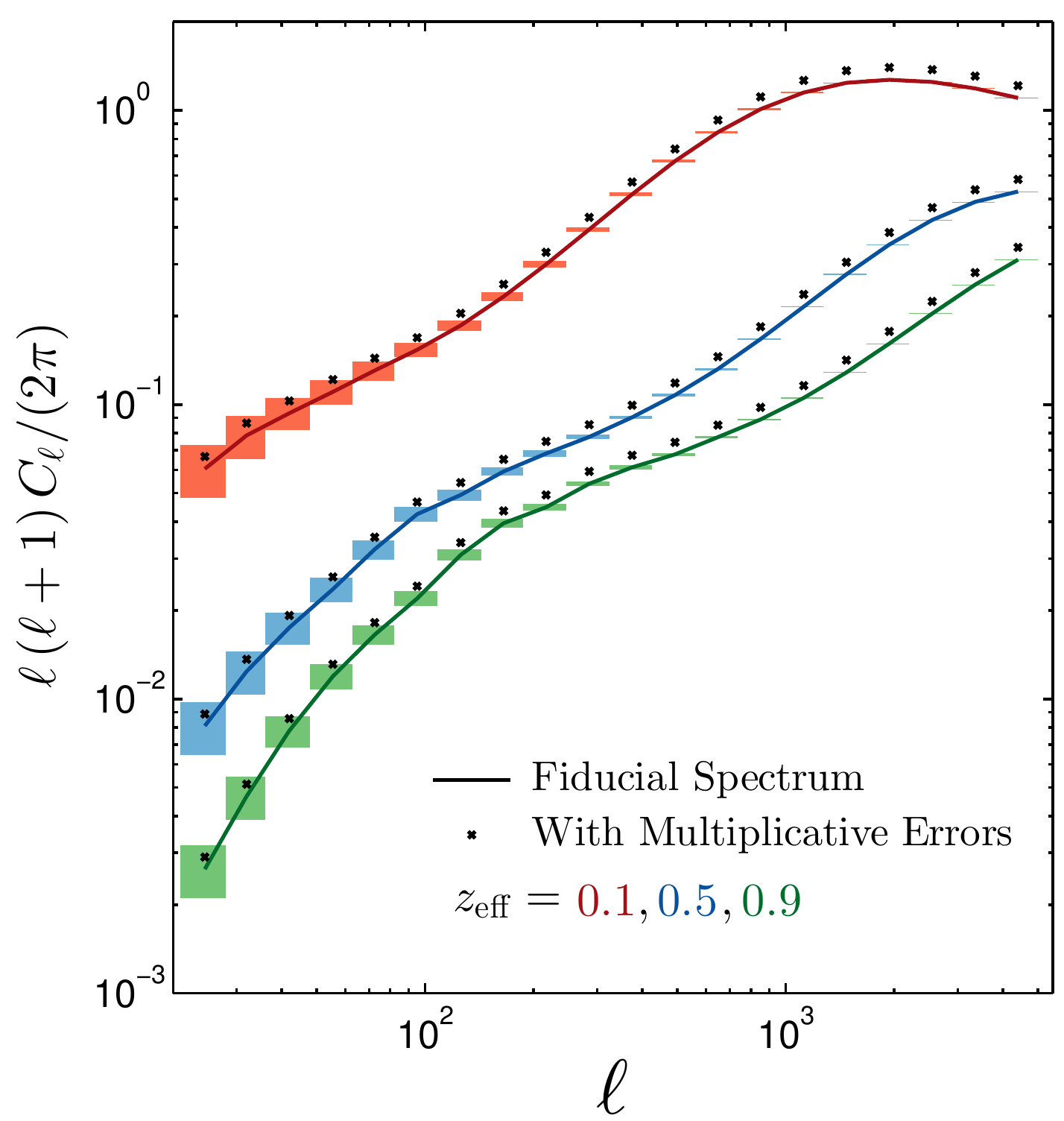}
\caption{Multiplicative effect due to our fiducial calibration errors with $\sigma^2_c = 0.1$ distributed on large scales $\ell \leq 20$. The biased power spectrum $T_\ell$ (black points) is compared to the true power spectrum $C_\ell$ (solid lines) for three of the five redshift bins of our fiducial survey. The spectra are binned in $\ell$ with inverse-variance weights, and the error boxes include cosmic variance and shot noise.}
\label{fig:Clbiased}
\end{figure}

\section{Results} \label{sec:results}

\subsection{Biases from multiplicative errors} \label{sec:bias}

H13 used a similar formalism to study the effect of arbitrary photometric calibration errors on cosmological parameters. The large biases that resulted were primarily due to additive errors (the $C_\ell^\text{cal}$ term in Eq.~\eqref{eq:Tl}), which strongly biased the power spectrum at low multipoles and which were assumed negligible at smaller scales ($C_\ell^\text{cal} = 0$ for $\ell > \ell_\text{max, cal}$). With the exception of some parameters (e.g.\ $f_\text{NL}$), most of the constraining power on cosmological parameters, including dark energy parameters, comes from high $\ell$, where there are many more modes to minimize cosmic variance. The simplest way to avoid biases due to these large additive errors at low $\ell$ is to just remove those multipoles from the analysis, sacrificing the modest amount of information they contain. For our fiducial survey, we find that this increases the errors on parameters by $\sim$1\%, though this number is sensitive to the model and which parameters are varied. Alternatively, with detailed modelling of the systematic effects, one can attempt to remove the contaminated modes and obtain useful cosmological information from the low multipoles.

The problem is not so simple when there are significant \emph{multiplicative} errors, corresponding to the other terms in Eq.~\eqref{eq:Tl}, where a given multipole is not only affected by calibration errors at that scale, but also calibration errors from every other multipole. Ignoring or cleaning the low-$\ell$ information is not helpful here, since the biases have already ``leaked'' into the high-$\ell$ power. The multiplicative errors are much smaller (by a factor of order $C_\ell$) than the additive errors, but for a large total amount of contamination, the effects can be important.

In Fig.~\ref{fig:Clbiased}, we show our fiducial power spectra along with the same power spectra when biased due to multiplicative errors, for three of the five redshift bins. The spectra are binned with inverse-variance weights, and the error boxes represent the combined cosmic variance and shot noise error for each bin. For definitiveness, we assume a spectrum for the calibration systematics of $C_\ell^\text{cal} \propto \ell^{-2}$ (separately for each redshift bin) and impose a cutoff such that $C_\ell^\text{cal} = 0$ for $\ell > \ell_\text{max, cal} = 20$. A variance of the calibration-error field across the sky of $\sigma^2_c = 0.1$ is assumed. Recall that this quantity is related to the power spectrum of calibration errors by $\sigma^2_c \equiv \text{Var}[c(\uv{n})] = \sum_{\ell = 1}^\infty (2\ell + 1) C_\ell^\text{cal}/(4\pi)$. While this is a large contamination, it may not be unrealistic for a survey like DES, since the relevant error is the raw variation in the effective magnitude limit before any attempts to clean or remove it. While the cleaning or marginalization methods effectively remove the additive contribution, the $C_\ell^\text{cal}$ term in Eq.~\eqref{eq:Tl}, the original multiplicative effects remain.

Note that the \emph{factor} by which the power is increased by the multiplicative errors is relatively constant, but since cosmic variance decreases at higher $\ell$, the biases eventually become larger than the errors. From the approximate expression Eq.~\eqref{eq:Tlapprox}, we can easily estimate this relative bias in the space of observables. The multiplicative bias is $\sigma^2_c \, C_\ell$, while the variance is given by Eq.~\eqref{eq:cosmicvar}. Ignoring the shot noise contribution, the bias relative to the error is therefore
\begin{equation}
\frac{T_\ell - C_\ell}{\sigma_{C_\ell}} \simeq \sqrt{\frac{(2\ell + 1) \, f_\text{sky}}{2}} \: \sigma^2_c
\end{equation}
For $\sigma^2_c = 0.1$, the bias is as large as the error for $\ell \simeq 800$ and twice as large for $\ell \simeq \text{3,000}$. The biases on bandpowers, such as those shown in Fig.~\ref{fig:Clbiased}, are more severe still, as cosmic variance is further reduced by measuring the power spectrum in bins spanning several independent $\ell$ modes.

\begin{figure}
\includegraphics[width=0.45\textwidth]{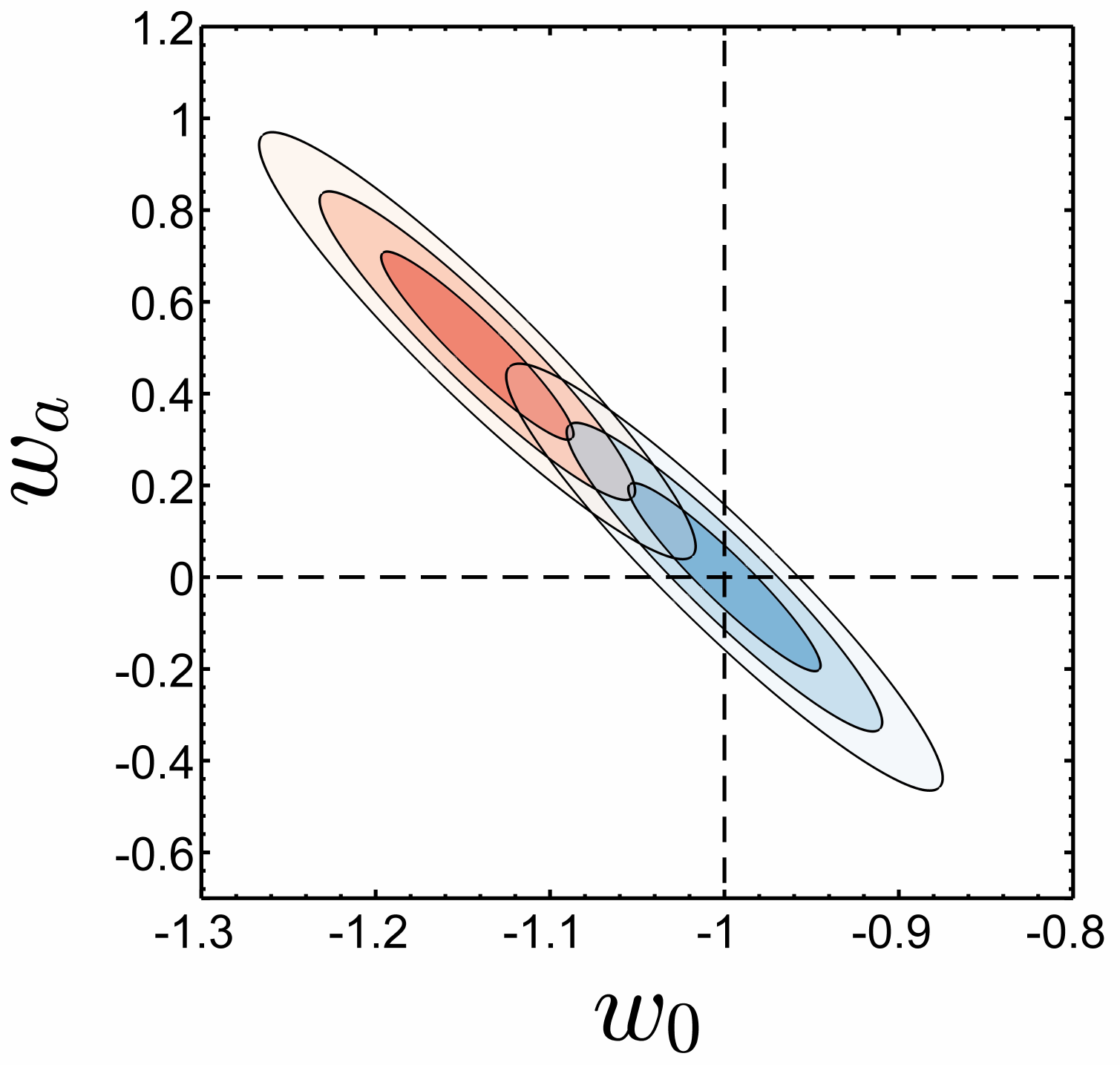}
\caption{Forecasted 68.3, 95.4, and 99.7 per cent joint constraints on the $w_0$--$w_a$ dark energy parametrization for our fiducial survey, using information from $\ell = 21$ through $\ell_\text{max} = \text{2,000}$ without calibration errors (blue) and with multiplicative calibration errors from $\ell \leq 20$ with $\sigma^2_c = 0.1$ (red).}
\label{fig:w0wa}
\end{figure}

Fig.~\ref{fig:w0wa} shows the effect of the bias from Fig.~\ref{fig:Clbiased} in the space of dark energy parameters $w_0$ and $w_a$ using information from $\ell = 21$ through $\ell_\text{max} = \text{2,000}$. In this case, both $w_0$ and $w_a$ are shifted from their fiducial values by more than 3$\sigma$.

\begin{figure}
\includegraphics[width=0.45\textwidth]{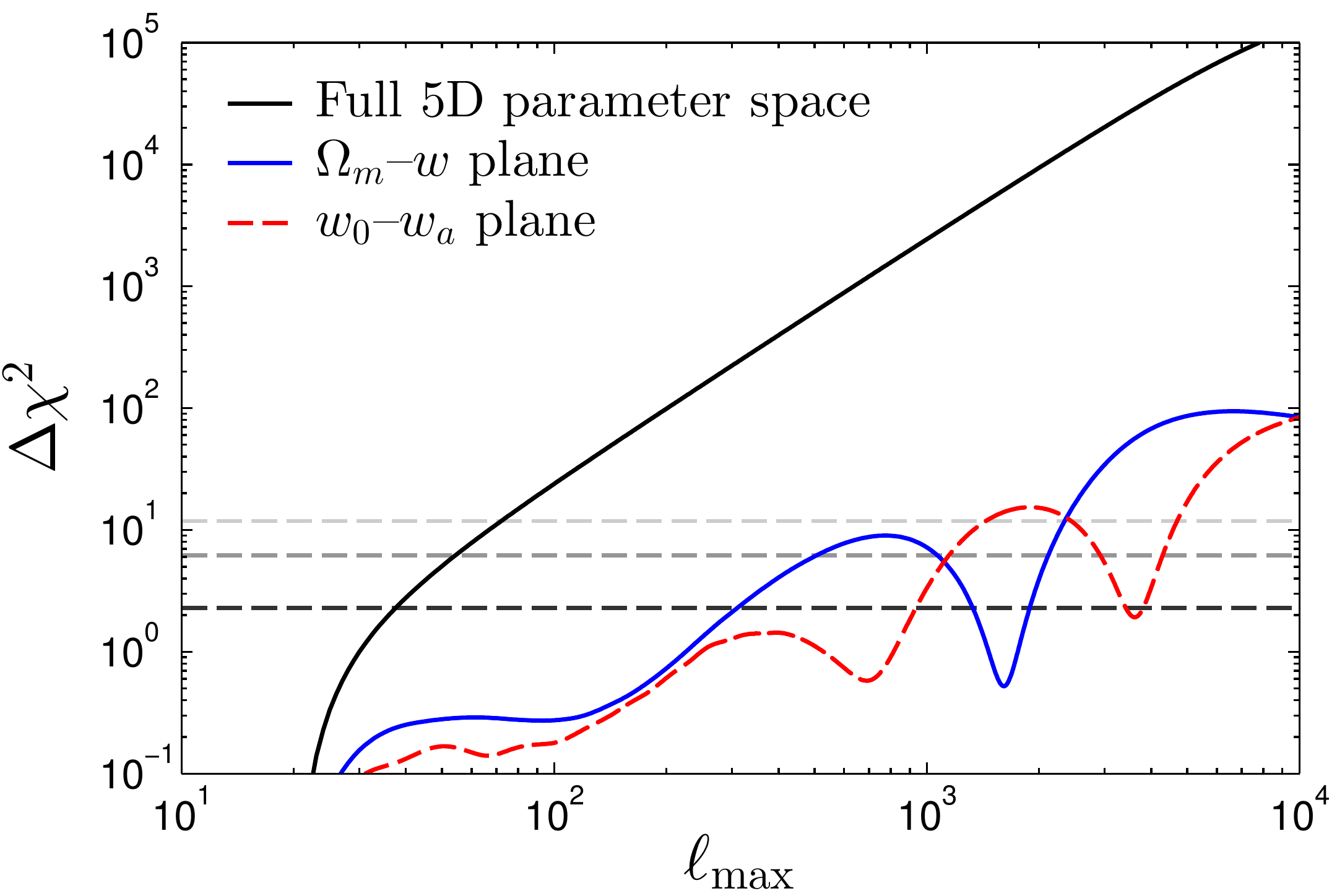}
\caption{Shift in parameter-space $\chi^2$ due to multiplicative calibration errors as a function of the maximum multipole used in the analysis. We show the effect on the full five-dimensional space of parameters (black) along with the two-dimensional spaces of $\Omega_m$ and $w$ with fixed $w_a = 0$ (blue) and $w_0$ and $w_a$ (red). The overlaid dashed grey lines mark the 68.3, 95.4, and 99.7 per cent bounds for a \emph{two-dimensional} Gaussian distribution (for comparison with the red or blue lines).}
\label{fig:dchisq}
\end{figure}

Fig.~\ref{fig:dchisq} shows the effect of these same biases in the full space of our cosmological parameters. We plot $\Delta \chi^2$ as a function of the maximum multipole $\ell_\text{max}$ used in the analysis, where $\Delta \chi^2 = \delta \mathbf{p}^\top \mathbf{F} \, \delta \mathbf{p}$. For the five-dimensional space of all parameters, this is equivalent to the observable-space $\Delta \chi^2$ in the Fisher matrix formalism. In this case, $\chi^2$ is shifted by $3 \sigma$ for $\ell_\text{max} \simeq 100$. We also show $\Delta \chi^2$ for the two-parameter spaces of $\Omega_m$ and $w$ (marginalizing over $A_s$ and $n_s$ but fixing $w_a = 0$) and for $w_0$ and $w_a$ (marginalizing over the other three parameters). In these cases, $\chi^2$ is shifted by more than $3 \sigma$ for $\ell_\text{max} \simeq \text{2,500}$. Notice that the sizes of the biases oscillate somewhat; since the Fisher derivatives sometimes flip sign, the biases will cancel for some $\ell_\text{max}$. This subtlety depends strongly on which parameters are of interest, apparent here from the ``out-of-phase'' cancelling between the constant-$w$ and $w_0$--$w_a$ dark energy parametrizations. It is therefore the envelope of the bias curves in the $\Omega_m$--$w$ and $w_0$--$w_a$ spaces that indicates the bias one can realistically expect.

Although we fix the galaxy bias in this illustration, it is worth mentioning that the effect at high $\ell$ of a constant (scale-independent) galaxy bias is almost completely degenerate with the multiplicative effect from low $\ell$ (see Eq.~\eqref{eq:Tlapprox} with $C_\ell^\text{cal} = 0$). In other words, if the contaminated low multipoles are removed from the analysis and the galaxy bias is constrained along with other cosmological parameters, the inferred value of the galaxy bias will shift due to the multiplicative effect, but the marginalized constraints on the other cosmological parameters will not be significantly biased. Of course, if the galaxy bias exhibits scale dependence or can be known independently to a good precision (which we effectively assumed here), this will not be the case.

\subsection{Self-calibration to remove multiplicative errors} \label{sec:selfcal}

We now study the possibility of measuring the contamination directly at low multipoles to correct the power at high multipoles. Due to large cosmic variance, the low-$\ell$ $C_\ell^\text{cal}$ are not known precisely, and for a cut sky, there will be very few modes to inform us about the lowest-$\ell$ $C_\ell^\text{cal}$. In practice, the $T_\ell$ would be measured in bandpowers, so one could then measure the calibration error in bandpowers, but for our purposes here, we assume that each $C_\ell^\text{cal}$ can be measured with an associated error due to cosmic variance of the true power (and shot noise, though it is negligible at the relevant low multipoles). Note that in our Fisher formalism we are ignoring any additional errors that may result from imperfectly extracting $T_\ell$ from the cut sky, though these could be estimated in principle \citep{Efstathiou:2003tv, Pontzen:2010yw, Leistedt:2013gfa}.

To study the effect of a self-calibration procedure, we consider the example in Sec.~\ref{sec:bias}, where a calibration power spectrum $C_\ell^\text{cal} \propto \ell^{-2}$ with $\sigma^2_c = 0.1$ and $\ell_\text{max, cal} = 20$ has been added to each of the fiducial power spectra. We introduce the $C_\ell^\text{cal}$ as nuisance parameters to be constrained along with the cosmological parameters. Since $\ell_\text{max, cal} = 20$, there are 20 calibration-error parameters for each of five redshift bins, for a total of 100 nuisance parameters. Using the Fisher matrix formalism discussed in Sec.~\ref{sec:fisher}, we can estimate the additional statistical error on cosmological parameters that results from imperfectly measuring the $C_\ell^\text{cal}$.

\begin{figure}
\includegraphics[width=0.45\textwidth]{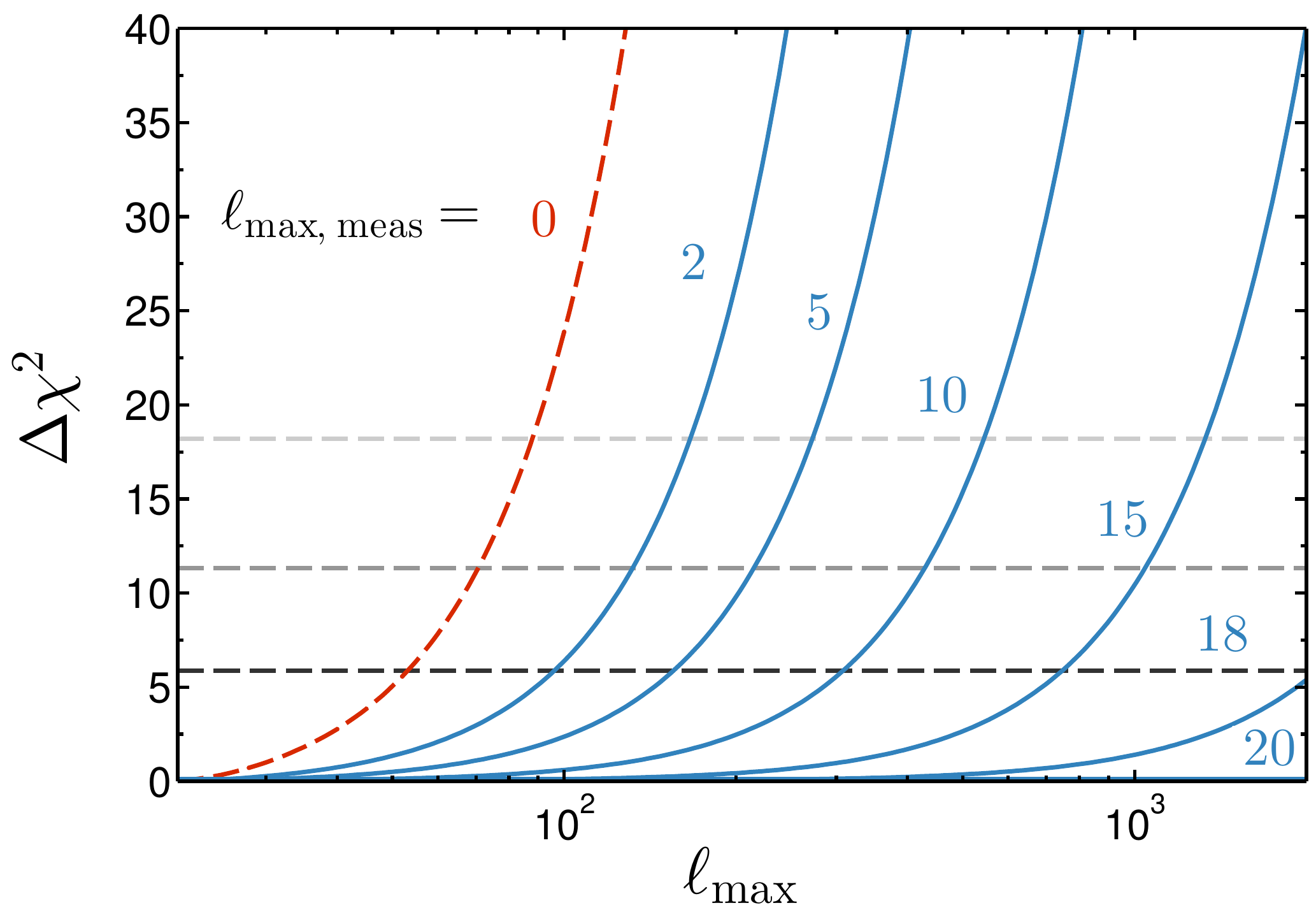}
\caption{Shift in the full five-dimensional parameter-space $\chi^2$ due to multiplicative calibration errors as a function of the maximum multipole used in the analysis, for calibration-error parameters measured up to various $\ell_\text{max, meas}$. The overlaid dashed grey lines mark the 68.3, 95.4, and 99.7 per cent bounds for a five-dimensional Gaussian distribution.}
\label{fig:dchisq_meas}
\end{figure}

In Fig.~\ref{fig:dchisq_meas}, we show the effect of measuring $C_\ell^\text{cal}$ up to a variety of $\ell_\text{max, meas}$ by plotting $\Delta \chi^2$ (in the five-dimensional parameter space of $\Omega_m$, $w_0$, $w_a$, $n_s$, and $A_s$) due to the remaining bias, as a function of the maximum multipole $\ell_\text{max}$ used in the analysis. In other words, for $\ell_\text{max, meas} = x$, we constrain $5 x$ total nuisance parameters. For $\ell_\text{max, meas} = 20$, $\Delta \chi^2 = 0$ for all $\ell_\text{max}$, since the assumption is that all of the calibration terms have been measured without significant bias. Note the very large biases in the cosmological parameters when $\ell_\text{max, meas}$ is low; that is, when we have not used measurements of additive error at sufficiently many low multipoles to effectively ``clean'' the high-multipole LSS information.

\begin{figure*}
\includegraphics[width=0.45\textwidth]{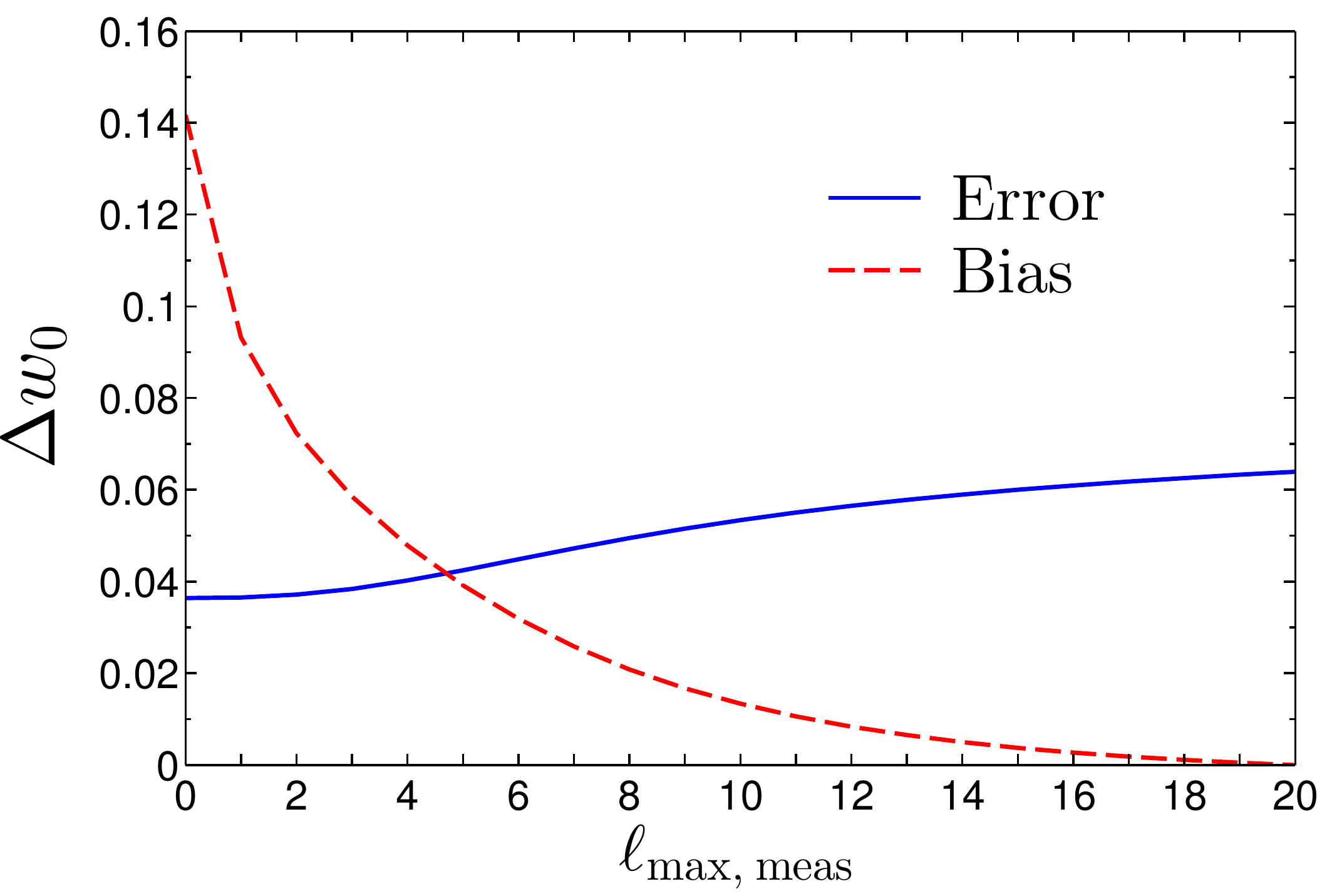}
\includegraphics[width=0.44\textwidth]{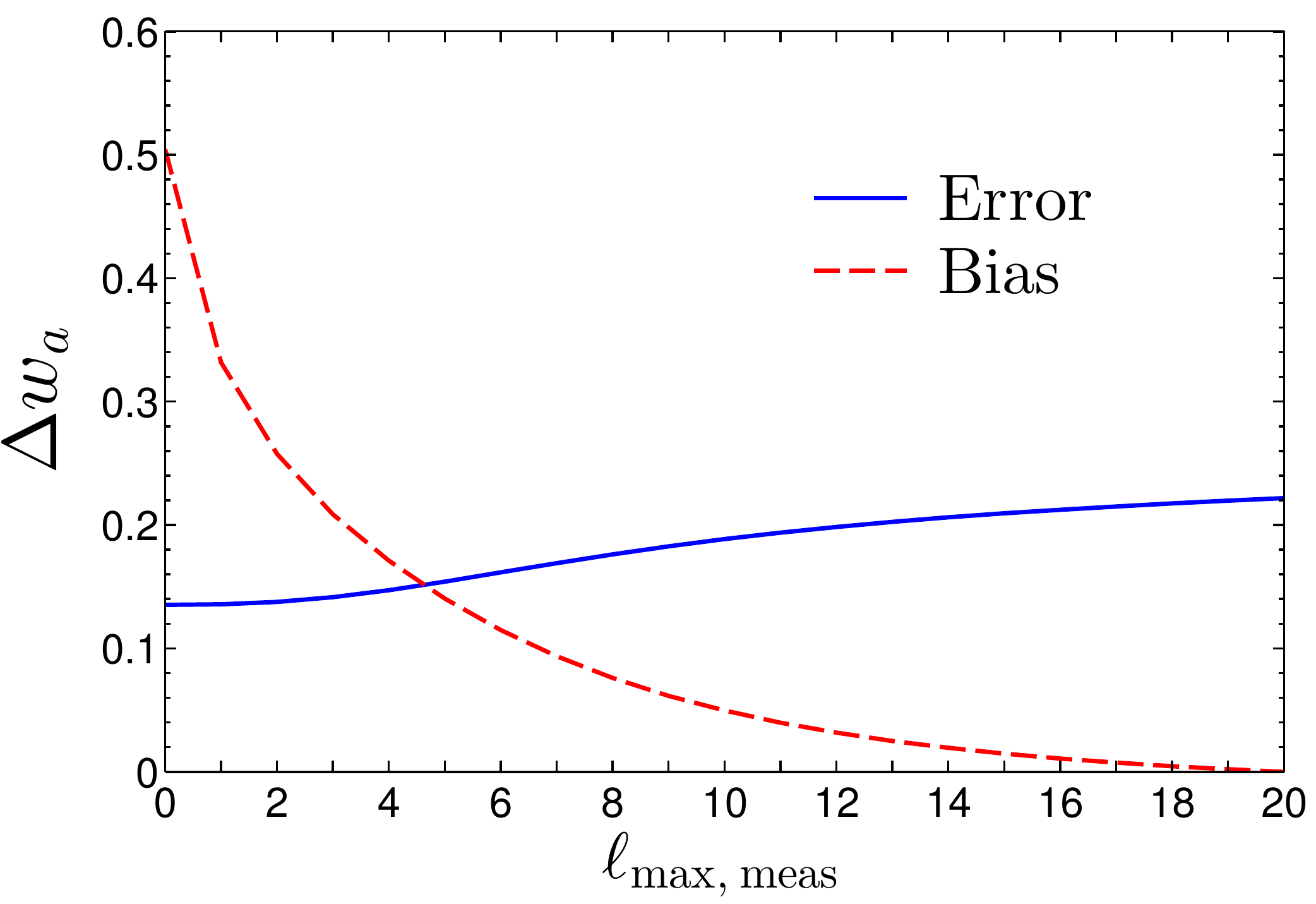}
\caption{Statistical error (blue) and bias (dashed red) on $w_0$ (left) and $w_a$ (right) as a function of the maximum multipole $\ell_\text{max, meas}$ at which calibration errors are measured.}
\label{fig:selfcal}
\end{figure*}

In Fig.~\ref{fig:selfcal}, we show the statistical error and remaining bias in the $w_0$ and $w_a$ dark energy parameters as a function of the maximum multipole $\ell_\text{max, meas}$ at which calibration errors are measured. For both parameters, the statistical errors increase modestly (by ${\sim}50\%$), while the biases approach zero at $\ell_\text{max, meas} = \ell_\text{max, cal}$. In this specific case, it is apparent that one would need to measure systematics to $\ell \simeq 10$ in order to reduce the biases to a comfortable level (such as ${\sim}1/4$ of the statistical error).

\section{Discussion} \label{sec:discuss}

In this paper, we have considered a general class of systematic errors -- photometric calibration errors -- that contaminate measurements of the galaxy angular power spectrum. These errors arise from any effect which causes a spatial variation in the effective magnitude limit of the photometric survey, modulating the true galaxy number densities and biasing the angular power spectrum (see Eq.~\eqref{eq:Tl}). More specifically, we studied the effect of \emph{multiplicative} errors, where calibration error at any scale biases inferences of the power spectrum at all other scales. In this case, cleaning the power spectra of excess additive power, or excluding contaminated multipoles from the analysis, does not remove the multiplicative effect. For a large total amount of contamination, the multiplicative effect can significantly bias cosmological parameters (Figs.~\ref{fig:w0wa}--\ref{fig:dchisq}).

Given the fact that these calibration errors tend to affect primarily large angular scales (for instance $\ell \lesssim 20$), we proposed a possible method of ``self-calibrating'' the survey by using the largest angular scales to measure the contamination itself, which can then be removed from small scales where most of the information on cosmological parameters resides. We studied a fiducial DES-like survey, using Fisher matrix formalism to forecast errors and biases in cosmological parameters given the survey parameters and our assumed photometric calibration error. We then extended the Fisher matrix to include the low-multipole calibration powers as nuisance parameters. For a modest increase in statistical uncertainty, one can remove the biases in cosmological parameters, including those describing dark energy (Fig.~\ref{fig:selfcal}).

We now briefly discuss how this method could be applied to real data. One clear problem with the procedure is the near-perfect degeneracy at large scales between the additive calibration-error power and cosmological parameters. This means that nearly all of the information on cosmological parameters must come from small scales. While little information on most cosmological parameters comes from the largest scales, a major exception is the non-Gaussianity parameter $f_\text{NL}$, on which \textit{most} information comes from these scales (see fig.~A1 of H13), making its measurement difficult with this procedure. On the other hand, it may be possible to incorporate cross-power spectra from overlapping redshift bins and cross-correlations with other probes into this formalism. The extra information, which may not be subject to the same systematics, could be used to constrain parameters like $f_\text{NL}$ along with the general calibration-error parameters.

The other important assumption is that one can safely ignore all additive error above some multipole $\ell_\text{max, cal}$. For our illustration here, we assumed $\ell_\text{max, cal} = 20$, though this is optimistic, and one could easily make the cutoff somewhere else. While photometric calibration error from known systematics tends to decrease sharply at smaller scales, some contamination may still be present at higher multipoles. Although one can probably assume that the extra multiplicative effect from any of these (smaller) additive errors is negligible, the additive errors themselves would need to be removed via other means.

One could thus imagine using this procedure in conjunction with mode projection and other cleaning techniques, using the cleanest possible small-scale spectrum but still retaining the fully contaminated spectrum at large scales to constrain the calibration-error parameters. This procedure would be particularly useful if one is worried about unknown sources of calibration error but willing to assume that these add significant power to multipoles below some cutoff only.

Finally, if one is doubtful about removing all of the smaller-scale additive contamination using standard cleaning techniques, the only way forward would be to choose a more conservative $\ell_\text{max, cal}$ that is high enough for one to comfortably assume that the bias in the remaining small-scale spectrum is due only to the multiplicative effect from the larger scales.

\section*{Acknowledgements}

We thank Boris Leistedt, Hiranya Peiris, Nishant Agarwal, Eduardo Rozo, Shirley Ho, Ashley Ross, and the anonymous referee for useful comments and discussions. We are supported by DOE Grant no. DE-FG02-95ER40899 and NSF Grant no. AST-0807564. DH thanks the Aspen Center for Physics, supported by NSF Grant no. 1066293, for hospitality.

\bibliographystyle{mn2e}
\bibliography{selfcal}

\appendix
\begin{onecolumn}
\section{The Biased Angular Power Spectrum} \label{sec:derive}

The observed (biased) angular power spectrum for tomographic redshift bins $i$ and $j$ is given by
\begin{align}
T_\ell^{ij} &= \frac{\sum_m \langle t_{\ell m}^i t_{\ell m}^{j*} \rangle}{2\ell + 1} \ , \\
t_{\ell m}^i &= \frac{1}{1 + \epsilon^i} \left[a_{\ell m}^i + c_{\ell m}^i + \sum_{\substack{\ell_1 m_1 \\ \ell_2 m_2}} R^{\ell_1 \, \ell_2 \, \ell}_{m_1 m_2 m} \, c_{\ell_1 m_1}^i a_{\ell_2 m_2}^i - \sqrt{4\pi} \ \epsilon^i \ \delta_{\ell 0} \right] \ , \\
\epsilon^i &= \frac{c_{0 0}^i}{\sqrt{4\pi}} + \frac{1}{4\pi} \sum_{\ell m} c_{\ell m}^i a_{\ell m}^{i *} \ ,
\end{align}
where the $R$ coupling is defined in Eq.~\eqref{eq:R}. Since the monopole of the calibration field (or equivalently, the true mean galaxy density $\bar{N}$) is not measurable, we are free to specify a value, so we choose $c_{0 0}^i = (-1/\sqrt{4\pi}) \sum_{\ell m} c_{\ell m}^i a_{\ell m}^{i *}$ so that $\epsilon^i = 0$. Then
\begin{align}
T_\ell^{ij} = \frac{1}{2\ell + 1} \sum_m &\left\langle \left[(a_{\ell m}^i + c_{\ell m}^i) \, (a_{\ell m}^{j*} + c_{\ell m}^{j*}) + \sum_{\substack{\ell_1 m_1 \\ \ell_2 m_2}} R^{\ell_1 \, \ell_2 \, \ell}_{m_1 m_2 m} \left[(a_{\ell m}^i + c_{\ell m}^i) \, c_{\ell_1 m_1}^{j*} a_{\ell_2 m_2}^{j*} + (a_{\ell m}^{j*} + c_{\ell m}^{j*}) \, c_{\ell_1 m_1}^i a_{\ell_2 m_2}^i \right] \right.\right. \\
&\left.\left. + \sum_{\substack{\ell_1 m_1 \\ \ell_2 m_2}} \sum_{\substack{\ell_1' m_1' \\ \ell_2' m_2'}} R^{\ell_1 \, \ell_2 \, \ell}_{m_1 m_2 m} R^{\ell_1' \, \ell_2' \, \ell}_{m_1' m_2' m} \, c_{\ell_1 m_1}^i a_{\ell_2 m_2}^i \, c_{\ell_1' m_1'}^{j*} \, a_{\ell_2' m_2'}^{j*} \right] \right\rangle \nonumber
\end{align}
Calculating the ensemble averages, we assume that the cosmological three-point function vanishes and that the $c_{\ell m}$ are fixed (not random) variables, with the exception of $c_{0 0}$ which must be considered separately. Using the definition of $C_\ell^\text{cal}$ in Eq.~\eqref{eq:Clcal}, we have (for $\ell \neq 0$)
\begin{align}
T_\ell^{ij} &= C_\ell^{ij} + C_\ell^{\text{cal}(ij)} - \frac{1}{4\pi} C_\ell^{\text{cal}(ij)} \left[C_\ell^{ii} + C_\ell^{jj} \right] + \frac{1}{4\pi} \sum_{\substack{\ell_1 \neq 0 \\ \ell_2 \neq 0}} (2\ell_1 + 1) \, C_{\ell_1}^{\text{cal}(ij)} \wj{\ell_1}{\ell_2}{\ell}{0}{0}{0}^2 (2\ell_2 + 1) \, C_{\ell_2}^{ij} \\
&+ \frac{1}{(4\pi)^2} \left[C_\ell^{\text{cal}(ij)} \left[C_\ell^{ii} C_\ell^{jj} + (C_\ell^{ij})^2 \right] + C_\ell^{ij} \sum_{\ell' \neq 0} (2\ell' + 1) \, C_{\ell'}^{ij} C_{\ell'}^{\text{cal}(ij)} \right] \ . \nonumber
\end{align}
Note that $a_{0 0}$, $C_0$, $t_{0 0}$, and $T_0$ are all equal to zero by construction, while $C_0^\text{cal}$ does not contribute and is left undefined. Restricting to auto-power spectra only, dropping the redundant redshift bin indices, and neglecting the last group of terms (which is suppressed by an extra factor of order $C_\ell$ relative to the other terms), we have
\begin{equation}
T_\ell = C_\ell + C_\ell^\text{cal} - \frac{1}{2\pi} C_\ell^\text{cal} C_\ell + \frac{1}{4\pi} \sum_{\substack{\ell_1 \neq 0 \\ \ell_2 \neq 0}} (2\ell_1 + 1) \, C_{\ell_1}^\text{cal} \wj{\ell_1}{\ell_2}{\ell}{0}{0}{0}^2 (2\ell_2 + 1) \, C_{\ell_2} \ ,
\end{equation}
which matches Eq.~\eqref{eq:Tl} in the text.

The following relations involving Wigner~3-$j$ symbols were useful for simplifying the expression for $T_\ell$ and for computing the symbols numerically:
\begin{align}
&\sum_m (-1)^m \wj{\ell}{\ell}{L}{m}{-m}{0} = (-1)^\ell \sqrt{2\ell + 1} \: \delta_{L 0} \label{eq:w3j1} \\
&\sum_{\ell m} (2\ell + 1) \wj{\ell_1}{\ell_2}{\ell}{m_1}{m_2}{m} \wj{\ell_1}{\ell_2}{\ell}{m_1'}{m_2'}{m} = \delta_{m_1 m_1'} \, \delta_{m_2 m_2'} \label{eq:w3j2} \\
&\sum_{m_1 m_2} (2\ell + 1) \wj{\ell_1}{\ell_2}{\ell}{m_1}{m_2}{m} \wj{\ell_1}{\ell_2}{\ell'}{m_1}{m_2}{m'} = \delta_{\ell \ell'} \, \delta_{m m'} \label{eq:w3j3} \\
&\wj{\ell}{\ell}{0}{m}{-m}{0} = \frac{(-1)^{\ell - m}}{\sqrt{2\ell + 1}} \label{eq:w3j4} \\
&\wj{\ell_1}{\ell_2}{\ell}{0}{0}{0} = (-1)^g \sqrt{\frac{(2 g - 2 \ell_1)!~(2 g - 2 \ell_2)!~(2 g - 2 \ell)!}{(2 g + 1)!}} \frac{g!}{(g - \ell_1)!~(g - \ell_2)!~(g - \ell)!} \quad \text{for integer} \ \ g = \frac{\ell_1 + \ell_2 + \ell}{2} \label{eq:w3j5}
\end{align}

\end{onecolumn}
\label{lastpage}
\end{document}